\documentclass[12pt]{article}

\usepackage{graphicx}
\usepackage{amsfonts}
\usepackage[top=2.75cm, bottom=2.75cm, left=2.75cm, right=2.75cm]{geometry} 

\newtheorem{proposition}{Proposition}

\newtheorem{definition}[proposition]{Definition}

\newtheorem{theorem}[proposition]{Theorem}

\def\+{{+\!\!\!+}}

\def\d{\partial} 
 
\def\a{\alpha}

\def\l{\lambda}

\def\F{\Psi} 
\def\e{\varepsilon}

\def\th{\theta}

\def\pmb#1{\setbox0=\hbox{#1}%
\kern.0em\copy0\kern-\wd0 
\kern-.04em\copy0\kern-\wd0 
\kern.08em\copy0\kern-\wd0 
\kern-.04em\raise.0433em\box0 } 
 
\def\half{\textstyle{1\over 2}}

\newcommand{\nc}{\newcommand} 
\nc{\beq}{\begin{equation}} 
\nc{\eeq}[1]{\label{#1}\end{equation}} 
\nc{\ber}{\begin{eqnarray}} 
\nc{\eer}[1]{\label{#1}\end{eqnarray}} 
\nc{\pek}[1]{\cite{#1}} 
\nc{\enr}[1]{(\ref{#1})} 
\nc{\kal}[1]{{\cal{#1}}} 
\nc{\dott}{\;\cdot\;} 
\nc{\coker}{\mathrm{coker}}
\nc{\ie}{{\it i.e.}}
\nc{\eg}{{\it e.g.}}
\newcommand{\Section}[1]{\section{#1} \setcounter{equation}{0}}

\def\0 {\nonumber}

\begin{document} 
\setcounter{page}{0}
\newcommand{\inv}[1]{{#1}^{-1}} 
\renewcommand{\theequation}{\thesection.\arabic{equation}} 
\newcommand{\be}{\begin{equation}} 
\newcommand{\ee}{\end{equation}} 
\newcommand{\bea}{\begin{eqnarray}} 
\newcommand{\eea}{\end{eqnarray}} 
\newcommand{\re}[1]{(\ref{#1})} 
\newcommand{\qv}{\quad ,} 
\newcommand{\qp}{\quad .} 

\thispagestyle{empty}
\begin{flushright} \small
UUITP-22/06 \\ HIP-2006-55/TH \\ YITP-SB-06-58 \\
\end{flushright}
\smallskip
\begin{center} \LARGE
{\bf Linearizing Generalized K\"ahler Geometry}
\\[12mm] \normalsize
{\bf Ulf~Lindstr\"om$^{a,b}$, Martin Ro\v cek$^{c}$, 
Rikard von Unge$^{d}$, and Maxim Zabzine$^{a}$} \\[8mm]
{\small\it
$^a$Department of Theoretical Physics 
Uppsala University, \\ Box 803, SE-751 08 Uppsala, Sweden \\
~\\
$^b$HIP-Helsinki Institute of Physics, University of Helsinki,\\
P.O. Box 64 FIN-00014 Suomi-Finland\\
~\\
$^c$C.N.Yang Institute for Theoretical Physics, Stony Brook University, \\
Stony Brook, NY 11794-3840,USA\\
~\\
$^{d}$Institute for Theoretical Physics, 
Masaryk University, \\ 61137 Brno, Czech Republic 
\\~\\
}
\end{center}
\vspace{10mm}
\centerline{\bfseries Abstract} \bigskip
The geometry of the target space of an $N=(2,2)$
supersymmetry sigma-model carries a generalized K\"ahler
structure. There always exists a real function, the
generalized K\"ahler potential $K$, that encodes all the
relevant local differential geometry data: the metric, the
$B$-field, {\it etc}. Generically this data is given by {\it
nonlinear} functions of the second derivatives of $K$. We
show that, at least locally, the nonlinearity on any
generalized K\"ahler manifold can be explained as arising
from a quotient of a space without this nonlinearity.
 \noindent 

\eject
\normalsize



\section{Introduction}

This paper is an elaboration and continuation of our
previous paper \cite{Lindstrom:2005zr}, where we proved the
existence of the generalized K\"ahler potential $K$ and
solved the problem of $N=(2,2)$ off-shell supersymmetry for
sigma-models. For a generic generalized K\"ahler manifold,
all geometrical data such as the metric $g$, the $B$-field
and the complex structures are expressible in terms of
derivatives of $K$. In the generic case $K$ enters
nonlinearily. In the present article we show that this
nonlinearity arises from a quotient construction. Indeed,
locally, any generalized K\"ahler manifold may be thought of
as a quotient of another generalized K\"ahler manifold with
very special properties.

At the level of the $N=(2,2)$ sigma-models our idea is
simple: If the model contains only (anti)chiral and twisted
(anti)chiral superfields, the left and right complex
structures commute, and {\em all geometrical data on the
target space is expressed linearly in terms of the
generalized K\"ahler potential}. We call such target spaces
bihermitian local product spaces (BiLPs). All nonlinearity
is related to the semichiral fields. We show that by gauging
certain symmetries, an appropriate combination of chiral and
twisted chiral fields may be traded for semichiral fields.
Thus any $N=(2,2)$ sigma-model can be understood as a
quotient of an $N=(2,2)$ sigma-model containing only chiral
and twisted chiral fields, \ie, defined over a specific BiLP
space (the auxiliary local product or ALP space). In this
paper we describe the properties of the ALP space and
present the details of the quotient. We attempt to present
the natural sigma-model construction in geometrical terms.

Mathematically we can formulate our results as follows:
Locally, for any generalized K\"ahler manifold $M$ we
consider an auxiliary space $M \times {\mathbb C}^{2d_s}$,
where $2d_s$ is the dimension of the cokernel of the
commutator of the left and right complex structures on $M$.
This auxiliary space is also a generalized K\"ahler manifold
but with the additional property that the two complex
structures commute. The fiber ${\mathbb C}^{2d_s}$ is
spanned by those vectors that respect the underlying
generalized K\"ahler geometry. All structures on $M$ can be
understood via the quotient construction described in this
paper.

The paper is organized as follows. In Section \ref{GKG} we
review the basic definitions and properties of generalized
K\"ahler geometry. In particular we discuss the special
class of these geometries where the generalized K\"ahler
potential enters linearly. Section \ref{OFF} recasts the
geometry in terms of $N=(2,2)$ sigma-models and sets up the
notation for further discussion. Section \ref{linear}
presents the central idea of this paper at the level of
$N=(2,2)$ sigma-models, namely, we show that any model with
semichiral fields can be thought of as a quotient of a model
that contains only chiral and twisted chiral fields. The
remaining Sections are devoted to the geometrical
explanation of this statement. Section \ref{ALPS} describes
ALPs (auxiliary local product spaces), which are built over
any generalized Kahler manifold as a principal bundle with a
free action of abelian isometries. Section \ref{quotients}
presents various quotient constructions used in the context
of sigma-models; in particular, we explain how to perform
the quotient on the ALPs. Section \ref{global} contains some
speculations about the possibility of a global version of
the quotient. Section \ref{summary} includes a summary of
the results and some open problems. We also include three
appendices in which we discuss possible potentials on
generalized K\"ahler manifolds, properties of the ALP space
generalized K\"ahler potential, and the explicit
calculations of the quotient data.

\section{Generalized K\"ahler geometry}
\label{GKG}

In this section, we present some geometrical background; we mostly
review well-known facts, but present some new observations as well.

\subsection{General definition and properties}
\label{generalities}

The definition of (twisted) generalized K\"ahler geometry
originates in the study of the general $N=(2,2)$
supersymmetric sigma-models \cite{Gates:1984nk}, although
the name and the renewed interest is due to recent work of
Gualtieri \cite{gualtieri}.

We define (twisted) generalized K\"ahler
geometry\footnote{Motivated by generalized complex geometry,
typically the word ``twisted" is added when $H\neq 0$. Here
we omit ``twisted" and assume that $H\neq 0$ unless
otherwise stated.} $(M, J_+, J_-, g, H)$ as the following
data on a smooth manifold $M$: $J_\pm$ are two complex
structures and $g$ is a metric which is bihermitian
\beq
J_\pm^t g J_\pm = g~.
\eeq{bla}
Moreover the complex structures $J_\pm$ are covariantly constant 
\beq
\nabla^{(\pm)} J_{\pm} = 0
\eeq{nablJH}
with respect to the connections with torsion
\beq
\Gamma^{\pm} = \Gamma \pm g^{-1} H~,
\eeq{defaffcon}
where $\Gamma$ is the Levi-Civita connection and $H$ is a
closed three-from. The name ``generalized K\"ahler geometry"
is motivated by the fact that when $J_+ = \pm J_-$ we
recover standard K\"ahler geometry.

Alternatively, we can define a generalized K\"ahler geometry
$(M, J_+, J_-, g)$ as two complex structures $J_\pm$ with a
bihermitian metric $g$ (\ref{bla}) and the integrability
conditions
\beq
d_+^c \omega_+ + d_-^c \omega_- =0~,~~~~~
d d^c_\pm \omega_\pm=0~,
\eeq{inegrajdsk3999}
where $\omega_\pm \equiv g J_\pm$ and $d_\pm^c$
are the $i(\bar{\d}- \d)$ operators associated to the 
complex structures $J_\pm$. The corresponding closed 
three-form that gives the torsion in the connections
(\ref{defaffcon}) is defined as
\beq
H = d_+^c \omega_+ = - d_-^c \omega_- ~,
\eeq{imdowjj390}
and is not an independent geometrical datum.

The generalized K\"ahler manifold $(M, J_+, J_-, g)$ admits
three different Poisson structures: two real Poisson
structures $\pi_\pm = (J_+ \pm J_-)g^{-1}$
\cite{Lyakhovich:2002kc} and the holomorphic Poisson
structure $\sigma= [J_+, J_-]g^{-1}$ \cite{hitchinP} with
the following obvious relation between the kernels
\beq
\ker \sigma = \ker \pi_+ \oplus \ker \pi_- ~.
\eeq{relationkernes}
We call $x_0 \in M$ a regular point of the generalized
K\"ahler manifold if the ranks of $\pi_\pm$ do not vary in
an open neighborhood of $x_0$. All other points of $M$ are
called singular points. The set of regular points of $M$ is
obviously open and is dense in $M$. If all points of $M$ are
regular we call such an $M$ a regular generalized K\"ahler
manifold.
\begin{theorem}
A generalized K\"ahler manifold with $H=0$ is regular.
\end{theorem}
{\it Proof:~} This follows immediately from Theorem $2.20$
in \cite{vaisman} and the fact that when $H=0$, the Poisson
structures $\pi_\pm$ are covariantly constant with respect
to the Levi-Civita connection. $\Box$

In the general case, in a neighborhood of a regular point we
can introduce coordinates along the symplectic foliations of
$\pi_\pm$ and $\sigma$; see \cite{Lindstrom:2005zr} for more
details. These local coordinates are adapted to the
following decomposition
\beq
\ker (J_+ - J_-) \oplus \ker (J_+ + J_-) \oplus \coker [J_+, J_-]~, 
\eeq{adpatedcoord}
where (the real) $\dim (\coker [J_+, J_-] )$ is a multiple
of four, as it corresponds to a symplectic leaf of the
holomorphic Poisson structure $\sigma$, and 
$\dim (\ker(J_+\pm J_-))$ is a multiple of two.

The generalized K\"ahler geometry can be nicely described in
the context of the generalized geometric structures
introduced by Hitchin \cite{hitchinCY}: a generalized
K\"ahler geometry is a pair of commuting (twisted)
generalized complex structures ${\cal J}_{1,2}$ such that
their product induces a definite metric on $TM\oplus T^*M$.
The equivalence of this definition with the one presented
before was proven in \cite{gualtieri}; see
\cite{Bredthauer:2006hf} for an alternative explanation of
this result via sigma-models. The following relation holds
between the type of (twisted) generalized complex structures
(see \cite{gualtieri} for the definition) and kernels of
$\pi_\pm$,
\beq
{\sf type} ({\cal J}_{1,2}) = \frac{1}{2} \dim (\ker \pi_\pm) 
= \frac{1}{2} \dim (\ker (J_+\pm J_-))~.
\eeq{defiofow99393}
Therefore a regular generalized K\"ahler manifold can be
defined as one where the type of ${\cal J}_{1,2}$ is
constant, \ie, no change in type occurs, and hence all
untwisted  generalized K\"ahler manifolds with $H=0$ are regular
(see the theorem above).

\subsection{The generalized K\"ahler potential} 
\label{kahlerpot}

In this subsection we review the arguments from
\cite{Lindstrom:2005zr} concerning the existence of a
generalized K\"ahler potential. We also comment on
symmetries of $K$.

Consider a neighborhood of a regular point $x_0$ of a
generalized K\"ahler manifold and choose local coordinates
adapted to the symplectic foliation of $\sigma$. We can
choose coordinates $\{q, p, z, z'\}$ in which $J_+$ has the
canonical form
\beq
J_+ = \left(\begin{array}{cccc}
J_s & 0 & 0& 0 \\
0 & J_s & 0 & 0\\
0 & 0 & J_c & 0\\
0 & 0 & 0 & J_t
\end{array} \right)~,
\eeq{canoniicalfromforJ}
where a collective notation is used in the matrices, and
where $J_ {c}, J_{t}$, and $J_{s}$ are $2d_{c}, 2d_{t}$, and
$2d_{s}$ dimensional canonical complex structures of the
form
\beq
J = \left( \begin{array}{ll}
i & ~\,0\\
0 & -i
\end{array}\right)~.
\eeq{defincanoncsysm}
The coordinates $z$ and $z'$ parametrize the kernels of
$\pi_\mp$ respectively, and $\{ q, p\}$ are the Darboux
coordinates for a symplectic leaf of $\sigma$. The subscript
``$c$" (chiral) corresponds to the coordinates along the
kernel of $\pi_-$, the subscript ``$t$" (twisted chiral)
corresponds to the coordinates along the kernel of $\pi_+$
and ``$s$" (semichiral) denotes the coordinates along the
leaf of $\sigma$.

Alternatively we can choose the coordinates $\{Q, P, z,
z'\}$ in which $J_-$ has a canonical form
\beq
J_- = \left(\begin{array}{cccc}
J_s & 0 & 0& 0 \\
0 & J_s & 0 & 0\\
0 & 0 & J_c & 0\\
0 & 0 & 0 & - J_t
\end{array} \right)~.
\eeq{canoniicalfromforJextra}
Again $(Q, P)$ are the Darboux coordinates on a leaf of
$\sigma$.

The coordinates $\{q, p\}$ are related to $\{Q, P\}$ by a
canonical transformation and we may thus introduce the
corresponding generating function. Choosing new coordinates
$\{ q, P\}$ along a leaf in a neighborhood of $x_0$, there
exists a family of generating functions $K(q, P, z, z')$
such that
\beq
p =\frac{\d K}{\d q}~,\qquad Q=\frac 
{\d K}{\d P}
\eeq{impiroao3994}
is satisfied. So far $K(q, P, z, z')$ is defined up to the
addition of an arbitrary function $F(z, z')$, which will be
partially fixed later on. We can also shift $K$ by a
$J_+$-holomorphic function $f(q,z, z')$ plus its complex
conjugate function, as this just gives a $J_+$-holomorphic
redefinition of the complex coordinates $\{ q, p, z, z'\}$.
Analogously we can shift $K$ by a $J_-$-holomorphic function
$g(P, z, z')$ plus its complex conjugate; this preserves
$J_-$. Thus these additional shifts are also natural
symmetries of our problem.

Moreover, since $K$ is a generating function, it is natural
to consider its Legendre transforms. For example, if we
would like to switch from $\{q, P, z, z'\}$ to $\{p, Q, z,z'\}$, 
then the corresponding generating function
$\tilde{K}(p, Q, z, z')$ is a Legendre
transform\footnote{This Legendre transform exists if $\{p,
Q, z, z'\}$ are good coordinates around $x_0$.} of $K(q, P,z, z')$
\beq
\tilde{K} = K-pq - QP 
\eeq{Legdenwq83299}
with (\ref{impiroao3994}) taken into account. Similar 
constructions relate the other possible generating 
functions. 

Next, using (\ref{impiroao3994}), we can perform a 
coordinate transformations from $\{q, p, z,z'\}$
to $\{q, P, z, z'\}$ and calculate $J_+$
\beq
J_+ = \left( \begin{array}{cccc}
J_s & 0 & 0 & 0\\
K^{-1}_{RL}C_{LL} & K_{RL}^{-1}J_s K_{LR} 
& K_{RL}^{-1}C_{Lc} &
K_{RL}^{-1}C_{Lt} \\
0 & 0 & J_c & 0 \\
0 &0 & 0 & J_t
\end{array} \right)~.
\eeq{fullJallcasesl}
The subscript ``$L$" stands for $q$, subscript 
``$R$" stands for $P$, the subscript $c$ for $z$ and 
the subscript $t$ is for $z'$. 
$K_{LR}$, $C_{LL}$ and $A_{LL}$, {\it etc.}, are included in 
the set of submatrices defined by
\ber
&K_{LR} \equiv\left(\begin{array}{cc}
K_{ab'} & K_{a\bar{b}'}\cr
K_{\bar{a}b'}& K_{\bar{a} \bar{b}'}\end{array}\right) \quad
&K_{LL} \equiv\left(\begin{array}{cc}
K_{ab} & K_{a\bar{b}}\cr
K_{\bar{a}b}& K_{\bar{a} \bar{b}}\end{array}\right)~,
\eer{HM}
and similarily for $K_{RR}$, as well as 
\ber
C &\equiv& JK-KJ
=\left(\begin{array}{cc}
0 & 2i K\cr
-2i K & 0
\end{array}\right),\cr\cr\cr
A &\equiv& JK+KJ
=\left(\begin{array}{cc}
2i K & 0\cr
0 & -2i K
\end{array}\right),
\eer{commnotation}
where the different possible subscripts have been suppressed. 
In (\ref{HM}) we use the notation $K_{ab} \equiv \d_a \d_b K 
$, {\it etc.},
and define $K_{LR}^{-1}\equiv (K_{RL})^{-1}$.

Analogously,
using (\ref{impiroao3994}), we can perform the coordinate 
transformations from $\{Q, P, z,z'\}$
to $\{q, P, z, z'\}$ and calculate $J_-$. The result is
\beq
J_-= \left( \begin{array}{cccc}
K_{LR}^{-1}J_s K_{RL} & K_{LR}^{-1}C_{RR} 
& - K_{LR}^{-1}C_{Rc} &
K_{LR}^{-1}A_{Rt} \\
0 & -J_s & 0 & 0\\
0 & 0 & J_c & 0 \\
0 &0 & 0 & -J_t
\end{array} \right)~.
\eeq{fullJallcaseslextra}

We now turn to the remaining data in our model: the metric
$g$ and the closed three-form $H$. The latter may be locally
expressed in terms of a two-form $B$ such that $H=dB$. It is
then convenient to define $E\equiv \half (g+B)$.

If $\sigma\equiv[J_{+},J_{-}]g^{-1}$ is invertible,
$E=J_+J_-\sigma^{-1}$; more generally, equations
(\ref{inegrajdsk3999}) and (\ref{imdowjj390}) provide
differential equations for $g$ and $B$ which may be solved
knowing $J_{\pm}$ and remembering that $\omega_{\pm}\equiv
g J_{\pm}$. The solution may again be expressed in terms of
the submatrices introduced in (\ref{HM}) and
(\ref{commnotation}). The solution is \cite{Lindstrom:2005zr}
\ber
E_{LL} &=& C_{LL}K_{LR}^{-1}J_sK_{RL} \cr
E_{LR} &=& J_sK_{LR}J_s + C_{LL}K_{LR}^{-1}C_{RR} \cr
E_{Lc} &=& K_{Lc} + J_s K_{Lc} J_c + C_{LL}K_{LR}^{-1}C_{Rc}\cr
E_{Lt} &=& -K_{Lt} - J_s K_{Lt} J_t + C_{LL}K_{LR}^{-1}A_{Rt}\cr
E_{RL} &=& -K_{RL}J_s K_{LR}^{-1} J_s K_{RL}\cr
E_{RR} &=& -K_{RL}J_s K_{LR}^{-1} C_{RR}\cr
E_{Rc} &=& K_{Rc} - K_{RL}J_s K_{LR}^{-1} C_{Rc}\cr
E_{Rt} &=& -K_{Rt} - K_{RL}J_s K_{LR}^{-1} A_{Rt}\cr
E_{cL} &=& C_{cL}K_{LR}^{-1}J_s K_{RL}\cr
E_{cR} &=& J_c K_{cR} J_s + C_{cL}K_{LR}^{-1}C_{RR}\cr
E_{cc} &=& K_{cc}+J_c K_{cc} J_c + C_{cL}K_{LR}^{-1}C_{Rc}\cr
E_{ct} &=& -K_{ct}-J_c K_{ct}J_t + C_{cL}K_{LR}^{-1}A_{Rt}\cr
E_{tL} &=& C_{tL}K_{LR}^{-1}J_s K_{RL}\cr
E_{tR} &=& J_t K_{tR} J_s + C_{tL}K_{LR}^{-1}C_{RR}\cr
E_{tc} &=& K_{tc} + J_t K_{tc} J_c + C_{tL}K_{LR}^{-1}C_{Rc}\cr
E_{tt} &=& -K_{tt} - J_t K_{tt} J_t + C_{tL} K_{LR}^{-1} A_{Rt}
\eer{E}
Thus we have locally expressed the generalized K\"ahler 
geometry in terms of a single function $K$, 
the generalized K\"ahler potential. 
This potential is not uniquely defined, as follows 
from our previous discussion: 
It can be Legendre transformed and shifted by (the 
real part of) a $J_\pm$-holomorphic function.

\subsection{Bihermitian local product geometry}
\label{BiLP}

In this subsection we draw attention to a special subset of generalized 
K\"ahler manifolds. We refer to a generalized K\"ahler geometry with the additional property 
$[J_+, J_-]=0$ as a bihermitian local product (BiLP) geometry. For BiLP geometries,
 $g$ and $H$ are linear in the generalized K\"ahler potential.
 
Equivalently, in the context of generalized geometry, we can define BiLP geometry as 
a generalized K\"ahler geometry with the addditional condition
\beq
{\sf type} ({\cal J}_1) + {\sf type}({\cal J}_2) = \frac{1}{2} \dim M ~, 
\eeq{codnewoi9920}
where $\dim M$ is the real dimension of $M$. The BiLP manifolds are examples
of regular generalized K\"ahler manifolds.\footnote{A global example of a BiLP 
structure can be found on the group manifold $SU(2)\times U(1)$; 
see \cite{Rocek:1991az,Rocek:1991vk}.}

Many important properties of BiLP geometry were already pointed 
out in \cite{Gates:1984nk}. 
In particular, there exists a local product structure $\Pi=J_+ J_-$,
which induces a decomposition of $TM$ into $\pm 1$-eigenspaces. 
This decomposition into $\pm 1$-pieces carries over to the differential forms
\beq
\Omega (M) = \bigoplus\limits_{l+m=d} \Omega^{l,m}(M)~.
\eeq{defdkk299301;k}
Furthermore there is a compatible decomposition with 
respect to $J_+$, \ie, each $\pm 1$ piece 
decomposes into holomorphic and antiholomorphic pieces 
correspondingly. Thus we have
the following decomposition of the differential forms 
\beq
\Omega (M) = \bigoplus\limits_{p+q+n+r=d} 
\Omega^{p,q,n,r}(M)~.
\eeq{evenslwe9o90}
This decomposition gives rise to a decomposition 
of the exterior derivative 
\beq
d= \d_\phi + \bar{\d}_\phi + \d_\chi + \bar{\d}_\chi
\eeq{diffderam}
in terms of four mutually anti-commuting differentials. 
Thus $(\phi, \chi)$ are $J_+$-holomorphic coordinates, 
$z\equiv(\phi,\bar{\phi})$ parametrize 
$\ker(J_+ -J_-)$, and $z'\equiv(\chi, \bar{\chi})$ 
parametrize $\ker (J_+ + J_-)$.

Now we can solve the conditions (\ref{inegrajdsk3999}) locally. 
The condition (\ref{imdowjj390}) becomes
\beq
H = i ( \bar{\d}_\phi + \bar{\d}_\chi - \d_\phi - \d_\chi) \omega_+ =
-i (\bar{\d}_\phi + \d_\chi - \d_\phi - \bar{\d}_\chi ) \omega_- ~,
\eeq{deriabdkhhekpa}
which implies
\beq
\d_\phi (\omega_+ + \omega_-)=0~,~~~~
\bar{\d}_\phi (\omega_+ + \omega_-)=0~,~~~~
\d_\chi (\omega_+ - \omega_-)=0~,~~~~
\bar{\d}_\chi(\omega_+ - \omega_-)=0 ~.
\eeq{setofeqisl}
These equations can be solved locally in terms 
of two real functions $K_1(\phi, \bar{\phi}, \chi, 
\bar{\chi})$ and $K_2(\phi, \bar{\phi}, \chi, \bar{\chi})$
\beq
\omega_+ = i \d_\phi \bar{\d}_\phi K_1 + i \d_\chi \bar{\d}_\chi K_2~,~~~~~~~~~
\omega_- = i \d_\phi \bar{\d}_\phi K_1 - i \d_\chi \bar{\d}_\chi K_2 ~.
\eeq{solution28238}
However, from (\ref{deriabdkhhekpa}) and 
$dH=0$ these two functions are related by
\beq
\d_\phi \bar{\d}_\phi \d_\chi \bar{\d}_\chi K_1 + 
\d_\phi \bar{\d}_\phi \d_\chi \bar{\d}_\chi K_2 =0
\eeq{relahdiu39889}
and therefore their sum can be written in the form 
\beq
K_1 + K_2 = f(\phi, \bar{\phi}, \chi) + 
\bar{f}(\bar{\phi}, \phi, \bar{\chi}) + g(\phi, \chi, \bar{\chi}) +
\bar{g}(\bar{\phi}, \bar{\chi}, \chi)~. 
\eeq{diffrehajw9o0}
Thus we may define a single function $K$ such that
\beq
K_1 = K + g(\phi, \chi, \bar{\chi}) + 
\bar{g}(\bar{\phi}, \bar{\chi}, \chi),~~~~~~~~~
K_2 = - K + f(\phi, \bar{\phi}, \chi) + 
\bar{f}(\bar{\phi}, \phi, \bar{\chi})~.
\eeq{definitionofKK}
This implies that we may write $\omega_\pm$ as 
\beq
\omega_+ = i \d_\phi \bar{\d}_\phi K - i\d_\chi \bar{\d}_\chi K,~~~~~~~~~
\omega_- = i\d_\phi \bar{\d}_\phi K +i \d_\chi \bar{\d}_\chi K~.
\eeq{solution28238bla}
Using the canonical form $J_{\pm}$ we define 
the metric $g$ via
\beq
\omega_\pm = gJ_{\pm}~.
\eeq{solution28238blah}
Finally, $H$ is given by
\beq
H= -\bar{\d}_\chi \d_\phi \bar{\d}_\phi K +
\d_\chi \d_\phi \bar{\d}_\phi K +
\bar{\d}_\phi \d_\chi \bar{\d}_\chi K - i 
\d_\phi \d_\chi \bar{\d}_\chi K ~.
\eeq{definofHHH}
Locally we have $H=dB$ and thus a possible representation of $B$ is
\beq
B = \bar{\d}_\phi \d_\chi K - \bar{\d}_\chi \d_\phi K. 
\eeq{localBB}
Hence, we have shown that all main objects 
are expressed locally in terms of second and third 
derivatives of a single real function $K$. 
The function $K$ is well-defined modulo the addition
of a function 
\beq
\Lambda(\phi, \chi) + \bar{\Lambda}(\bar{\phi}, 
\bar{\chi}) + L(\phi, \bar{\chi}) +\bar{L}(\bar{\phi}, \chi)~.
\eeq{ BiLPkahlegana}
We would like to stress again that $K$ enters linearily 
in all formulas and that this is an
essential feature of the BiLP geometry. 

\section{$N=(2,2)$ supersymmetric sigma-models}
\label{OFF}

In this Section we translate the properties of 
generalized K\"ahler geometry into sigma-model 
language. We start from the general $N=(1,1)$ sigma-model 
written in terms of $N=(1,1)$ superfields
\beq
S = \int_\Sigma d^2\sigma\,d^2\theta~
D_+\Phi^\mu D_- \Phi^\nu E_{\mu\nu} 
(\Phi) ~,
\eeq{actionB}
where $E=\frac{1}{2}(g+B)$ with $H=dB$. 
Requiring the existence of an
additional supersymmetry transformations of the 
form \cite{Gates:1984nk}
\beq
\delta_2(\epsilon) \Phi^\mu
=\epsilon^+ D_+ \Phi^\nu J^\mu_{+\nu}(\Phi)
+ \epsilon^- D_- \Phi^\nu J^\mu_{-\nu}(\Phi)~,
\eeq{secsupfl}
we find that the target space must have generalized K\"ahler
geometry.

Next we introduce the $N=(2,2)$ superfields needed for a
complete description of the $N=(2,2)$ sigma-model in $(2,2)$
superspace. We work in the coordinates adapted to the
decomposition (\ref{adpatedcoord}):\\ $\bullet$ $\ker (J_+ -
J_-)$\\ These directions are parametrized by chiral $\phi$
and antichiral $\bar{\phi}$ fields defines as
\beq
\bar \mathbb{D}_{\pm}\phi=\mathbb{D} _{\pm}\bar\phi=0 ~.
\eeq{definchirall}\\
$\bullet$ $\ker(J_++J_-)$\\
These directions are parametrized by twisted 
chiral $\chi$ and twisted antichiral $\bar{\chi}$ fields
defined as
\beq
\bar\mathbb{D}_{+}\chi= \mathbb{D}_{-}\chi=
\mathbb{D}_{+}\bar\chi=\bar\mathbb{D}_{-}\bar \chi=0~.
\eeq{bla192920}\\
$\bullet$ $\coker [J_+, J_-]$\\
These directions are parametrized by 
left semichiral $\mathbb{X}_L$ and left 
anti-semichiral fields $\bar{\mathbb{X}}_L$
\beq
\bar \mathbb{D}_{+}\mathbb{X}_L= 
\mathbb{D}_+ \bar\mathbb{X}_L = 0~,
\eeq{left}
and right semichiral $\mathbb{X}_R$ and 
right anti-semichiral fields $\bar{\mathbb{X}}_R$
\beq
\bar\mathbb{D}_- \mathbb{X}_R = 
\mathbb{D}_{-}\bar{\mathbb{X}}_R = 0~.
\eeq{right}\\
Here $\mathbb{D}$ is the $N=(2,2)$ covariant 
derivative and we
follow the notation in \cite{Lindstrom:2005zr}. 
The chiral and twisted chiral superfields
were studied in \cite{Gates:1984nk}. The semichiral 
superfields were introduced in \cite{Buscher:1987uw}. 
In \cite{Lindstrom:2005zr} we have proven that these 
superfields provide a full description of the general 
$N=(2,2)$ sigma-model.
The most general sigma-model action is 
specified by a real function $K$ 
\beq
S = \int d^2\sigma d^4\theta~ K(\phi,\bar\phi,\chi,
\bar\chi,\mathbb{X}_L,\bar{\mathbb{X}}_L,
\mathbb{X}_R,\bar{\mathbb{X}}_R)~.
\eeq{Kgen}
From a geometrical point of view, the function
$K$ is precisely the generalized K\"ahler potential 
that we discussed in the previous section.

The potential $K$ is not uniquely defined: it 
can be shifted by the following combination
\beq
f(\phi, \chi,\mathbb{X}_L)+g(\phi,
\bar\chi,\mathbb{X}_R )+
\bar f(\bar\phi,\bar\chi,\bar\mathbb{X}_L)+\bar 
g(\bar\phi,\chi,\bar \mathbb{X}_R)~,
\eeq{PgeneralgaugeK}
which changes the action (\ref{Kgen}) at most by 
total derivatives. We refer to this shift as
a ``generalized K\"ahler gauge transformation''. 
The geometrical interpretation of (\ref{PgeneralgaugeK}) 
was given in the previous Section,
see the discussion after (\ref{impiroao3994}). We identify 
$(\mathbb{X}_L,\bar\mathbb{X}_L)$ with the coordinates 
$q$ and $(\mathbb{X}_R,\bar\mathbb{X}_R)$ 
with the coordinates $P$.

Furthermore, we can perform a Legendre transform along 
$\coker ([J_+, J_-])$, \ie, along the 
semichiral directions \cite{Grisaru:1997pg}. 
Starting from a parent action
\beq
\int d^2\sigma d^4\theta~ \left (K(\phi, \bar{\phi}, \chi, 
\bar{\chi}, U,\bar U,V,\bar V)-\mathbb{X}_L 
U-\bar\mathbb{X}_L\bar U-\mathbb{X}_R V-
\bar\mathbb{X}_R\bar V\right )~,
\eeq{LTR}
where $U$ and $V$ are unrestricted superfields,
we may choose to integrate out semichiral superfields 
$ \mathbb{X}_L$ and $\mathbb{X}_R$, which restricts 
$U=U_{L}$ to be left semichiral
and $V=V_{R}$ to be right semichiral. The resulting semichiral action is
$K(\phi, \bar{\phi}, \chi, \bar{\chi}, U_L,\bar U_L,V_R,\bar V_R)$.
Integrating out $U$ and $V$ instead, we solve the system
\ber
&&\mathbb{X}_L=K_{U}\quad \bar\mathbb{X}_L=K_{\bar U}\cr
&&\mathbb{X}_R=K_{V}\quad \bar\mathbb{X}_R=K_{\bar V}~,
\eer{Ltf}
to get $U=U(\phi, \bar{\phi}, \chi, \bar{\chi}, 
\mathbb{X}_L,\bar\mathbb{X}_L,\mathbb{X}_R,
\bar\mathbb{X}_R)$, {\it etc.}
Substituting the solution into \enr{LTR} 
yields the Legendre transformed action
$\tilde K(\phi, \bar{\phi}, \chi, \bar{\chi}, 
\mathbb{X}_L,\bar\mathbb{X}_L,\mathbb{X}_R,\bar\mathbb{X}_R)$.
Similarily, integrating out $ \mathbb{X}_L$ and $V$ yields a
third dual $K$ and integrating out $ \mathbb{X}_R$ and $U$
yields a fourth. These symmetries are directly related to
the fact that $K$ has an interpretation as a generating
function and therefore the Legendre transform corresponds to
switching to different generating function as described in
subsection \ref{kahlerpot}.

To extract $g$ and $B$ from (\ref{Kgen}) we need to
integrate out some of the Fermi-coordinates $\theta$ and get
rid of auxiliary fields and thus arrive at the $N=(1,1)$
action (\ref{actionB}). We introduce some notation which we
use throughout the rest of the paper. The superfields carry
indices of the following kind and range;
\ber
\phi^\alpha, \bar\phi^{\bar\alpha}~,~~\alpha=1\dots d_c&~,~~&
\chi^{\alpha'}, \bar\chi^{\bar \alpha'}~,~~\alpha ' =1\dots d_t~,~~\nonumber\\
\mathbb{X}_L^{a}, \bar\mathbb{X}_L^{\bar a}~,~~a =1\dots d_s&~,~~&
\mathbb{X}_R^{a'},\bar\mathbb{X}_R^{\bar a'}~,~~a'=1\dots d_s~.
\eer{indices}
We also use the collective notation $\kal{A}\equiv(\alpha, \bar 
\alpha)$,
$\kal{A}'\equiv(\alpha ', \bar\alpha ')$, $A\equiv(a, \bar
a)$ and $A'\equiv(a', \bar a')$. To reduce the $N=(2,2)$
action to its $N=(1,1)$ form, we introduce the $N=(1,1)$
superspace derivatives $D$ and extra supercharges $Q$
\cite{Gates:1984nk}:
\beq
D_{\pm}=\mathbb{D}_{\pm}+\bar \mathbb{D}_{\pm}~,~~~~
Q_{\pm}=i(\mathbb{D}_{\pm}-\bar \mathbb{D}_{\pm})~.
\eeq{twone}
In terms of these, the (anti)chiral, twisted (anti)chiral and 
semi (anti)chiral superfields satisfy
\ber
Q_{\pm}\phi = J_c D_{\pm} \phi &~,~~&
Q_\pm \chi = \pm J_t D_\pm \chi~,\cr
Q_+ \mathbb{X}_L= J_s D_+ \mathbb{X}_L&~,~~&
Q_- \mathbb{X}_R = J_s D_- \mathbb{X}_R~.
\eer{QDrels}

For the pair $(\phi,\chi)$
we use the same letters to denote the $N=(1,1)$
superfields, \ie, the lowest components of the $N=(2,2)$ superfields 
$(\phi,\chi)$.
Each of the semi (anti)chiral fields gives rise to two $N=(1,1)$ 
fields \cite{Buscher:1987uw}:
\ber
&&X_{L}\equiv \mathbb{X}_L|\qquad \F_{L-}\equiv Q_{-}\mathbb{X}_L |\cr
&&X_{R}\equiv \mathbb{X}_R|\qquad \F _{R+}\equiv Q_{+}\mathbb{X}_R |~,
\eer{twone2}
where a vertical bar means that we take the
$\theta^2\propto\theta-\bar\theta$ independent component.
The conditions \enr{QDrels} then also imply
\ber
&&Q_{+}\F_{L-}=J_{s}D_{+}\F_{L-}~,\qquad Q_{-}\F_{L-}= i\partial_ 
{=}X_{L}\cr
&&Q_{-}\F_{R+}= J_{s}D_{-}\F_{R+}~,\qquad Q_{+}\F_{R+}=
i\partial_{\+}X_{R}~.
\eer{QF}

Using the relations \enr{twone}-\enr{QF} we reduce the $N=(2,2)$
action to its $N=(1,1)$ form according to:
\beq
\int d^2\sigma d^2\th d^2\bar\th~ K (\phi^{\kal{A}}, \chi^{\kal{A}'}, 
\mathbb{X}_L^{A},\mathbb{X}_R^{A'})|
=\int d^2\sigma \mathbb{D} ^2\bar \mathbb{D}^2 K|=-\frac i 4 \int d^2 
\sigma D^{2}Q_{+}Q_{-}
K|~.
\eeq{Act2}
Provided that the matrix $K_{LR}$ (\ref{HM})
is invertible, the auxiliary spinors $\F_{L-},\F_{R+}$ may be integrated
out leaving us with a
$N=(1,1)$ second order
sigma-model action of the type originally discussed in \cite{Gates:1984nk}.
From this the metric and antisymmetric
$B$-field may be read off in terms of derivatives of $K$, and from the
form of the second supersymmetry (\ref{QDrels}-\ref{QF}) 
the complex structures $J_{\pm}$ are determined.
We shall use a basis where the coordinates are arranged in a column as
\ber
\left(\begin{array}{c}
X_{L}^{A}\cr
X_{R}^{A'}\cr
\phi^{\kal{A}}\cr
\chi^{\kal{A}'}\end{array}\right)~.
\eer{Column}
when we compute the $N=(1,1)$ Lagrangian; the 
sum $E=\frac12(g+B)$ of the metric $g$ and 
$B$-field then takes on the explicit form (\ref{E}) \cite{Sevrin:1996jr}.

It is interesting that there are no corrections from chiral and twisted
chiral fields in the semichiral sector (where the results agree with
\cite{Buscher:1987uw} and \cite{Lindstrom:2004hi}),
whereas in the chiral and twisted chiral sector the semichiral fields 
contribute substantially.

Thus, locally, all objects ($J_\pm$, $g$, $B$) are given in terms of 
second derivatives of a single real function $K$. By construction,
the present geometry is generalized K\"ahler,
and therefore satisfies all the relations from the previous section.

In addition we can add to (\ref{actionB}) the 
potential term $\int d^2\sigma d^2\theta~ W(\Phi)$
and ask about the most 
general $N=(2,2)$ Landau-Ginzburg model. 
We present the detailed analysis in Appendix A.
The upshot is that the only allowed terms one can add to (\ref{Kgen}) are 
\beq
\int d^2\sigma\,d\theta^+ d\theta^- {\cal W}(\phi)+ 
\int d^2\sigma\,d\theta^+ d\bar{\theta}^- \tilde{\cal W}(\chi) + c.c~.
\eeq{generalLGmodel}
and no potential is allowed in semichiral directions.

\section{Linearization}
\label{linear}

As we have stressed a few times, the general 
expression (\ref{E}) for $E$ is nonlinear. 
However, a superficial look at (\ref{E}) 
suggests that nonlinearity is of the quotient 
type. This is the main point of our paper: any generalized 
K\"ahler data comes from a quotient of a BiLP 
geometry with respect to a set of abelian isometries. 

We start by presenting an $N=(2,2)$ sigma-model argument. 
At this level the quotient idea is almost obvious. 
Consider the action
\beq
\int d^2 \sigma d^4\theta ~K(\phi, \bar{\phi}, 
\chi, \bar{\chi}, \phi_L + \chi_L, \bar{\phi}_L +
\bar{\chi}_L, \phi_R + \bar{\chi}_R, \bar{\phi}_R + \chi_R)~,
\eeq{gaughesoo2900}
where we have taken the action (\ref{Kgen}) and replaced the
semichiral entries by the combinations of new chiral and
twisted chiral fields: $\phi_L$, $\phi_R$, $\chi_L$ and
$\chi_R$. Thus the theory (\ref{gaughesoo2900}) is defined
over a manifold with a dimension 
$\left (\dim M + \dim(\coker[J_+, J_-]) \right )$. Since the action
(\ref{gaughesoo2900}) depends only on chiral and twisted
chiral superfields, the target geometry is of the BiLP type.
The action (\ref{gaughesoo2900}) has the following {\it
complex} symmetries
\beq
\delta \phi_L = \lambda_L~,~~~~\delta \chi_L = -\lambda_L~,~~~~
\delta \phi_R = \lambda_R~,~~~~\delta \chi_R = -\bar{\lambda}_R~.
\eeq{symmetrieso12}
The parameters satisfy 
$$
\bar\mathbb{D}_\pm \lambda_L=0~,~~~\mathbb{D}_-\lambda_L=0~,~~~
\bar\mathbb{D}_\pm \lambda_R=0~,~~~\mathbb{D}_+\lambda_R=0~,
$$
and thus they correspond to Kac-Moody symmetries, 
\ie, $\d_{=} \lambda_L=0$ and $\d_{++} \lambda_R=0$.

We can gauge these symmetries using semichiral fields as ``connections''.
The action 
\beq
\int d^2 \sigma d^4\theta ~K(\phi, \bar{\phi}, \chi, \bar{\chi}, 
\phi_L + \chi_L+\mathbb{X}_L, \bar{\phi}_L +
\bar{\chi}_L+\bar{\mathbb{X}}_L, \phi_R + 
\bar{\chi}_R+\mathbb{X}_R, \bar{\phi}_R + \chi_R+\bar{\mathbb{X}}_R)
\eeq{gaughesoo2900new}
is invariant under the gauge transformations
\ber
&& \delta \phi_L = \Lambda_L~,~~~~~\,
\delta \chi_L = - \tilde{\Lambda}_L~,~~~~~\,
\delta \mathbb{X}_L = -\Lambda_L + \tilde{\Lambda}_L~,\\
&& \delta \phi_R = \Lambda_R~,~~~~~\,
\delta \bar{\chi}_R = - \tilde{\Lambda}_R~,~~~~~\,
\delta \mathbb{X}_R = -\Lambda_R + \tilde{\Lambda}_R~,
\eer{symmetrusiool929}
(as well as their complex conjugates). 
The parameters $\Lambda_{L,R}$ 
are chiral whereas the parameters 
$\tilde{\Lambda}_{L,R}$ are twisted chiral. 
By fixing the gauge symmetry in (\ref{gaughesoo2900new}), 
\beq
\phi_L + \chi_L =0~,~~~~~~~~\phi_R + \bar{\chi}_R=0~,
\eeq{fixingsymmak002}
we arrive at the action (\ref{Kgen}). Thus an $N=(2,2)$
sigma-model (\ref{Kgen}) on a generalized K\"ahler geometry
can be thought of as a quotient of the sigma-model given by
(\ref{gaughesoo2900}) defined on a certain BiLP geometry. In
the rest of the paper we will translate this statement into
geometrical terms. We refer to the target space BiLP
geometry defined by (\ref{gaughesoo2900}) as an auxiliary
local product space (ALP space).
\section{Defining the ALPs}
\label{ALPS}

In this Section we describe the ALPs in intrinsic geometrical terms.
We are only concerned with the local properties of ALPs here, 
and our speculations about the global 
structure are presented in Section \ref{global}. 
In particular we have to understand the extra 
requirements a BiLP geometry must satisfy for it to be an 
ALP, \ie, to have an generalized K\"ahler 
potential form in (\ref{gaughesoo2900}). 

The generalized K\"ahler potential 
in (\ref{gaughesoo2900}) is invariant 
under the action of the following vector fields
\beq
k_L^a= \frac{\d}{\d \phi^a_L} - \frac{\d}{\d \chi^a_L}~,~~~~~~~~~~
k_R^{a'}= \frac{\d}{\d \phi^{a'}_R} - \frac{\d}{\d \bar{\chi}^{a'}_R}~,
\eeq{isometri2999}
as well as their complex conjugates. 
The invariance of $K$ with respect to these 
vectors implies that the metric $g$
and $H$ are invariant as well. Thus these vectors 
generate isometries. Moreover $K$ obeys 
\beq
K_{\phi_L^a \bar{\phi}_L^b} = K_{\chi_L^a \bar{\chi}_L^b}~,~~~~~~~~~~
K_{\phi_R^{a'}\bar{\phi}_R^{b'}} = K_{\chi_R^{a'}\bar{\chi}_R^{b'}}~;
\eeq{propermetrisoo}
these relations imply that the metric $g$ is indefinite. 
Thus $k_L$ and $k_R$ 
span two isotropic (null) subspaces with respect 
to the metric $g$ defined in 
(\ref{solution28238blah}). 
However, for the original model (\ref{Kgen})
to make sense, we must require that 
\beq
K_{\phi_{L}^a \bar{\phi}_R^{b'}}
\eeq{matricKRLRL}
is a nondegenerate matrix. Furthermore, the vectors $(\Pi
k_{L},\Pi k_R)$ are linearly independent of $(k_{L},k_R)$,
where $\Pi$ is the product structure of the BiLP geometry.

We now reformulate all these properties in a coordinate
independent way. The ALP space is locally a trivial
fibration over the generalized K\"ahler manifold $M$ with
fiber the vector space $\mathbb{C}^{2d_s}$, where $2d_s =
\dim (\coker [J^M_+, J^M_-])$ and the complex vector fields
$k_L$ and $k_R$ span $\mathbb{C}^{2d_s}$ (for clarity, we
indicate quantities on the original manifold with a
superscript $M$). Further, the vectors $k_L$ and $k_R$ in
(\ref{isometri2999}) correspond to the left and right
Kac-Moody symmetries (\ref{symmetrieso12}) of the
$\sigma$-model, and, as observed in \cite{Rocek:1991ps},
satisfy
\beq
\nabla^{(+)} k_{LA}=0~,~~~~~~~~~~\nabla^{(-)} k_{RA'}=0~,
\eeq{Pdefinusiapep}
where the corresponding connections are 
defined in (\ref{defaffcon}). 
We use the property (\ref{Pdefinusiapep})
as the {\it definition} of left (right) Kac-Moody 
isometries of the geometry.

As every $k_L$ is matched with a corresponding 
$k_R$, we identify the indices
$A'\simeq A$ below.
From (\ref{Pdefinusiapep}) it follows that 
\beq
{\cal L}_{k_L} g =0~,~~~~~~~
{\cal L}_{k_R} g =0~,~~~~~~~
{\cal L}_{k_L} H=0~,~~~~~~~
{\cal L}_{k_R} H =0~.
\eeq{isometrydefsh}
We also require that the isometries $k_L$ and $k_R$ 
respect the BiLP geometry; 
thus in addition to (\ref{isometrydefsh}) 
we have the conditions
\beq
{\cal L}_{k_L} J_\pm =0~,~~~~~~~~~~\,
{\cal L}_{k_R} J_\pm =0~. 
\eeq{commusysyai}
Finally the conditions (\ref{propermetrisoo}) become
\beq
k_{LA}^\mu g_{\mu\nu} k_{LB}^\nu =0~,~~~~~~~~~~\,
k_{RA}^\mu g_{\mu\nu} k_{RB}^\nu=0~,
\eeq{nullcondinvar}
whereas $k_{LA}^\mu g_{\mu\nu} k_{RB}^\nu$ is required to be
nondegenerate.

\begin{definition} We define the ALP geometry as a BiLP
geometry\footnote{Recall that a BiLP geometry is a
generalized K\"ahler geometry with the additional
requirement $[J_+, J_-]=0$, see the discussion in sec.
\ref{BiLP}.} with an equal number of left and right
Kac-Moody null abelian isometries (\ref{Pdefinusiapep}) that
respect the BiLP geometry and in addition satisfy
(\ref{nullcondinvar}). Furthermore, we require that the
product structure $\Pi$ maps the isometry directions into
$\coker[J_+^M, J_-^M]$. \end{definition}

Thus locally an ALP space has the structure $M \times (V_L
\oplus V_R)$ in addition to the BiLP geometry. Here $V_L$
($V_R$) is an isotropic vector space with respect to the
metric $g_{ALP}$ and it is spanned by $k_L$ ($k_R$) which
are left (right) Kac-Moody isometries. The product structure
$\Pi$ maps $(V_L \oplus V_R)$ to the vectors tangential to
$\coker[J_+, J_-]$ in $M$.

It is a straightforward exercise to show that $K$ in
(\ref{gaughesoo2900}) satisfies this definitions. Indeed,
this motivated the definition. However, the inverse
statement that any ALP $K$ can brought to the form
(\ref{gaughesoo2900}) is less trivial and we collect the
details of this proof in Appendix B.

\begin{theorem} Locally, any generalized K\"ahler manifold
$M$ with its geometrical data can be thought of as a
quotient of an ALP geometry by its Kac-Moody isometries.
\end{theorem} This is consistent with dimension of the ALP
and the number of isometires:
$$ \dim (ALP) = \dim M + \dim (\coker [J_+, J_-])~.$$

This theorem explains the nonlinearities that
appear in the metric and $B$ field when they are expressed
in terms of a generalized K\"ahler potential. 
The proof of this theorem is quite trivial at the level of 
sigma-model--see the previous Section. 
However, the geometrical aspects of this theorem are less trivial,
and the rest of the paper is devoted to clarifying these. 

\section{Quotients and sigma-models}
\label{quotients}

In this Section we review some old and discuss some new
aspects of quotients. In particular, we are interested in
the relation between the sigma-model approach to quotients
and their geometrical interpretation. Our main goal is to
explain how to perform the quotient with respect to
Kac-Moody null isometries. Everywhere we will assume that
the isometries commute.

We start by recalling some standard material on quotient
metrics and sigma-models; for more details see
\cite{Hitchin:1986ea}. We need this background to contrast
it with the more exotic quotient constructions that follow.
Consider a smooth manifold $\mathbb{M}$ and assume that a
Lie group ${\mathbf G}$ acts smoothly on $\mathbb{M}$. If
${\mathbf G}$ acts freely and properly then the quotient
space $\mathbb{M}/{\mathbf G}$ (\ie, the space of orbits) is
a smooth manifold. There is a corresponding principal bundle 
\beq
\begin{array}{ccc}
\mathbb{M}& \stackrel{}{\longleftarrow} &{\mathbf G}\\ 
{\scriptstyle p}\Big\downarrow &&\\
\mathbb{M}/{\mathbf G} &&
\end{array} 
\eeq{principalbundle}
with $p$ being a smooth projection. If $\mathbb{M}$ has a
metric $g$ and ${\mathbf G}$ acts as isometries, then we can
define a metric $\tilde{g}$ on the quotient space. The
corresponding nonzero vector fields $k_A$ generated by
${\mathbf G}$ form a basis for a vertical subspace $V_m$ of
$T_m{\mathbb M}$, for $m\in {\mathbb M}$. Defining the
horizontal subspace $H_m$ as the set of vectors orthogonal
to $V_m$ we can define the metric $\tilde{g}$ on the
quotient as
\beq
\tilde{g}(v, w) = g(\tilde{v}, \tilde{w})~,
\eeq{defisnwoooP} 
where $v, w \in T_{p(m)} (\mathbb{M}/{\mathbf G})$ and their
unique horizontal lift $\tilde{v}, \tilde{w} \in H_m \subset
T_m \mathbb{M}$. Since $\mathbf{G}$ preserves the metric
$g$, this definition of $\tilde{g}$ is independent of the
choice of point $m$ in the orbit $p^{-1}(p(m))$.

Alternatively, the choice of a horizontal subspaces can be
described in terms of a choice of connection on the
principal bundle (\ref{principalbundle}). The orthogonal
projection from $T_m \mathbb{M}$ to $V_m$ defines a one-form
$\theta$ with values in a Lie algebra. In the present
context the connection is defined as
\beq
\theta^A_\mu = H^{AB}k_B^\nu g_{\nu\mu}, 
\eeq{definconnennann}
where $H^{AB}$ is the inverse of the matrix $H_{AB}\equiv 
k_{A}^\mu g_{\mu\nu}k_{B}^\nu$.
Clearly,
\beq
\theta^A_\mu k^\mu_B = \delta^A_{~\,B}. 
\eeq{starsthhjjjjk}
Using the connection form $\theta$
in local coordinates, we get the expressions for $\tilde{g}$
\beq
\tilde{g}_{\mu\nu} 
= g_{\mu\nu} - g_{\mu\rho} k^\rho_A \theta^A_\nu 
= g_{\mu\nu} - \theta_\mu^A
k_A^\rho g_{\rho\nu} = g_{\mu\nu} - 
\theta^A_\mu H_{AB} \theta^B_\nu ~.
\eeq{localcoodkwell}
Clearly,
\beq
k^\mu_A \tilde{g}_{\mu\nu} =0~.
\eeq{defisd3090300}

The above geometrical picture for the quotient 
metric arises naturally in the sigma-model
framework: 
Consider the bosonic sigma-model\footnote{The 
supersymmetric $N=(1,1)$ sigma-models
are treated in an identical fashion.} defined over $\mathbb{M}$ 
\beq
S=\int d^2\sigma~ \d_{+}X^\mu g_{\mu\nu}\d_{-}X^\nu~.
\eeq{bal}
Since $k_A$ are the Killing vectors for $g$ there 
is a corresponding global symmetry of the action $S$, given by 
\beq
\delta X^\mu=\e^{A}k_{A}^\mu ~.
\eeq{isvar}
By introducing the world-sheet gauge fields $A$ 
we promote this global symmetry to a gauge symmetry
\beq
S_{gauge} = \int d^2\sigma~ (\d_{+}X^\mu + k_A^\mu A_{+}^A) g_{\mu\nu}
(\d_{-}X^\nu + k^\nu_B A_{-}^B)~.
\eeq{gauegasjkk222}
Extremizing the action $S_{gauge}$ with respect to $A$,
we obtain a relation between the world-sheet and 
target space connections
\beq
A^{A} =-\theta^A_\mu dX^\nu = - X^*(\theta^A). 
\eeq{minpot}
Thus the world-sheet gauge field $A$ is just a pull-back of the connection
$\theta$ by the map $X$. Reinserting this 
form of $A$ in the action $S_{gauge}$ yields the quotient sigma-model action,
which is naturally defined over the space of orbits, $\mathbb{M}/{\mathbf G}$
\beq
S=\int d^2\sigma~ \d_{+}X^\mu \tilde{g}_{\mu\nu}\d_{-}X^\nu~,
\eeq{quomet7338388}
where $\tilde{g}$ is defined in (\ref{localcoodkwell}). The sigma-model 
derivation of the quotient metric naturally produces the connection form $\theta$.

We now consider more exotic quotients. In particular, we assume 
that $\mathbb{M}$ admits an invariant metric $g$ and an 
invariant two-form\footnote{This property can be relaxed 
and one can require that the closed three-from $H=dB$ 
is invariant \cite{Hull:1989jk}. 
However, we do not discuss this general case here.}, \ie,
\beq
{\cal L}_{k_A} g =0~,~~~~~~~~~~
{\cal L}_{k_A} B = 0. 
\eeq{deribamdlllslaopp} 
The corresponding bosonic sigma-model on $\mathbb{M}$
\beq
S=\int d^2\sigma~ \d_{+}X^\mu E_{\mu\nu}\d_{-}X^\nu
\eeq{ba1e}
with $E=\frac12(g+B)$ is invariant under a 
global symmetry (\ref{isvar}). Gauging this symmetry 
exactly in the same fashion as before and 
extremizing the gauge action we find
\ber
&&A^{A}_{+}=- (\theta_L)^A_{\mu} \d_{+}X^\mu \cr
&&A^{A}_{-}=- (\theta_R)^A_{\mu} \d_{-}X^\mu ~,
\eer{minpot2}
with the target space connections are defined as
\beq
(\theta_L)^{A}_{\mu} = E_{\mu\nu}k_{B}^{\nu} {\cal H}^{BA},~~~~~~~~~
(\theta_R)^A_{\mu} = {\cal H}^{AB}k_{B}^{\nu}E_{\nu\mu} ~,
\eeq{p}
where ${\cal H}^{AB}$ is the inverse 
of ${\cal H}^{AB}:=k_{A}^\mu E_{\mu\nu}k_{B}^\nu$. 
Now we have two different connection 
forms $\theta_L$ and $\theta_R$, left and right,
which both satisfy
\beq
(\theta_L)^A_\mu k^\mu_B = \delta^A_{~\,B},~~~~~~~~\,
(\theta_R)^A_\mu k^\mu_B = \delta^A_{~\,B}. 
\eeq{conemmao87348}
Plugging (\ref{minpot2}) into the gauged 
action produces the quotient $\tilde{E}$ 
\beq
\tilde{E}_{\mu\nu}=E_{\mu\nu}- 
(\theta_L)_\mu^Ak^{\rho}_{A}E_{\rho \nu} =
E_{\mu\nu} - E_{\mu\rho}k^\rho_{A} 
(\theta_R)_\nu^A = E_{\mu\nu} - (\theta_L)_\mu^A {\cal H}_{AB}
(\theta_R)_\nu^B
\eeq{quote}
which satisfies
\beq
k^\mu_A \tilde{E}_{\mu\nu}=0~,~~~~~~~~~~~
\tilde{E}_{\mu\nu} k^\nu_A=0
\eeq{aodo399r8r8}
The quotient metric and B-field are given by the symmetric
and antisymmetric part of $\tilde E$, respectively. Thus we
see that the sigma-model offers a different quotient which
involves the choice of two different connections, \ie,
different choice of left and right horizontal spaces.

Finally, we discuss an even more exotic quotient involving
null Kac-Moody isometries. This is the quotient that
we actually use to descend from the ALP space to the underlying
generalized K\"ahler geometry. We derive this quotient 
using a sigma-model construction, and then describe it geometrically.

Consider a manifold $\mathbb{M}$ that admits two sets of
null abelian Kac-Moody isometries generated by left and
right vector fields $k_{LA}$ and $k_{RA}$, respectively.
Assume that these vector fields satisfy the left and right
Kac-Moody condition
\beq
\nabla^{(+)} k_{LA}=0~,~~~~~~~~~~\nabla^{(-)} k_{RA}=0~,
\eeq{KMcodnammm}
and moreover that they are null
\beq
k^\mu_{LA} g_{\mu\nu} k_{LA}^\nu =0~,~~~~~~~~~~\,
k^\mu_{RA} g_{\mu\nu} k_{RA}^\nu =0 ~.
\eeq{nullcondlklsla} 

%
The conditions (\ref{KMcodnammm}) imply invariance of the metric
as well as the conditions
\beq
k_{LA}^\mu H_{\mu\nu\rho}=\d_{[\nu}\a^{LA}_{\rho]}~,~~~~~~~
k_{RA}^\mu H_{\mu\nu\rho}=-\d_{[\nu}\a^{RA}_{\rho]}
\eeq{sil}
with $\alpha_\mu = g_{\mu\nu}k^\nu$.


Now consider a sigma-model on $\mathbb{M}$ with
the standard action (\ref{ba1e}), which chosen to be invariant
under the Kac-Moody symmetries
\beq
\delta X^\mu = \e_L^A (z) k^\mu_{LA} (X) + \e_R^A (\bar{z}) k^\mu_{RA} (X)~.
\eeq{invariamsdl23999}
Inspired by \cite{Hull:1989jk}, we gauge this symmetry 
(promoting $\e_L(z)\to\l_L(z,\bar z), ~\e_R(\bar z)\to\l_R(z,\bar z)$)
and obtain the gauged action
\beq
S_{gauge} = \int d^2\sigma~ (\d_{+}X^\mu + 
 k^\mu_{RA} \tilde{A}^A_+) g_{\mu\nu}
(\d_{-}X^\nu + k^\nu_{LB} A_{-}^B )
-\frac12 \partial_+X^\mu 
\left(g_{\mu\nu}-B_{\mu\nu}\right) \partial_- X^\nu~,
\eeq{gaugeactaiiopps}
with
\beq
\delta A = -d\l_L~,~~~~~\delta \tilde A =- d\l_R~.
\eeq{dAdAt}
Note that only half of the gauge fields appear in the 
gauged action (\ref{gaugeactaiiopps}): The fields $A_+,\tilde{A}_-$
does not appear, but nevertheless the action is fully gauge invariant. 
The various terms in the action are not gauge invariant by themselves 
but their variations cancel (after integration by parts). There is
also a term from the variation of the first term that is linear in gauge fields, which of
course could never be cancelled by the second term. Fortunately it is zero using
the null condition (\ref{nullcondlklsla}).

Extremizing this action gives
\ber
&&\tilde{A}^{A}_{+}=- (\theta_R)^A_{\mu} \d_{+}X^\mu \cr
&& A^{A}_{-}=- (\theta_L)^A_{\mu} \d_{-}X^\mu ~,
\eer{djdkdfkl299303}
with the target space connections defined as
\beq
(\theta_R)^A_\mu = g_{\mu\nu}  k^{\nu}_{LB}h^{BA},~~~~~~~~~
(\theta_L)^A_\mu = h^{AB}k^\nu_{RB}  g_{\nu\mu}~,
\eeq{definsalxldoo}
where $h^{AB}$ is the inverse of 
\beq
h_{AB} = k_{RA}^\mu g_{\mu\nu} k_{LB}^\nu~,
\eeq{Pdjewke99930-0}
and must be nondegenerate for the construction 
to work. The connections (\ref{definsalxldoo})
satisfy the following properties
\beq
(\theta_L)^A_\mu k^\mu_{LB}=\delta^A_{~\,B},~~~~~\,
(\theta_R)^A_\mu k^\mu_{RB}=\delta^A_{~\,B},~~~~~\,
(\theta_L)^A_\mu k^\mu_{RB}= 0~,~~~~~\,
(\theta_R)^A_\mu k^\mu_{LB}= 0. 
\eeq{definsll2000}
Finally the quotient sigma model gives rise to $\tilde{E}$
\beq
\tilde{E}_{\mu\nu} = E_{\mu\nu} - E_{\mu\rho} 
k^\rho_{LA} (\theta_L)^A_\nu =
E_{\mu\nu} - (\theta_R)^A_\mu k^\rho_{RA} 
E_{\rho\nu} = E_{\mu\nu} - (\theta_R)^A_\mu h_{AB}
(\theta_L)^B_\nu ~.
\eeq{djkdkwe000}
This gives $\tilde{g}$ which satisfies 
\beq
k_{RA}^\mu \tilde{g}_{\mu\nu}=0~,~~~~~
\tilde{g}_{\mu\nu} k^\nu_{LA}=0~,~~~~~
k_{LA}^\mu\tilde{g}_{\mu\nu}=0~,~~~~~
\tilde{g}_{\mu\nu} k_{RA}^\nu =0
\eeq{stausiooooo}
and thus it is a well-defined tensor on the quotient. 
The $\tilde{B}$ we get from this $\tilde{E}$ is not zero when
contracted with any killing vector. However, when we compute
$\tilde{H}$ from $\tilde{B}$ using the identities
\beq
k_{LA}^\mu k_{LB}^\nu H_{\mu\nu\rho} = 0~,~~
k_{RA}^\mu k_{RB}^\nu H_{\mu\nu\rho} = 0~,~~
k_{LA}^\mu k_{RB}^\nu H_{\mu\nu\rho} = \frac12 \partial_\rho h_{BA}~,
\eeq{bident}
we get
\beq
\tilde{H}_{\mu\nu\rho} = H_{\mu\nu\rho}
- k_{LA}^{\alpha}(\theta_L)^{A}_{[\mu}H_{\nu\rho ] \alpha}
-k_{RA}^{\alpha}(\theta_R)^{A}_{[\mu} H_{\nu\rho]\alpha}
+k_{LA}^{\alpha} k_{RB}^\beta (\theta_L)^{A}_{[\mu}
(\theta_R)^{B}_{\nu} H_{\rho]\alpha\beta} ~,
\eeq{Hquot}
which obeys
\beq
\tilde{H}_{\mu\nu\rho} k_{LA}^{\rho}=\tilde{H}_{\mu\nu\rho} k_{RA}^{\rho}=0~.
\eeq{Htilk}


This is the geometric form
of the quotient that allows us to descend from
the ALPs to produce a generalized K\"ahler manifold. 

\section{Global Issues}
\label{global}

Hitherto, we have discussed only the local geometry of the ALP space. 
{\it A priori}, it is not clear if our construction 
makes sense globally, at least in the present form.
We would like to make a few speculative remarks 
about the possibility of a global interpretation. 

Consider diffeomorphisms along the leaves of the Poisson structure
$\sigma$ that preserve our special coordinates. 
In the sigma-model this corresponds to the following
\beq
\int d^2 \sigma d^4\theta ~K\left ( \phi, \bar{\phi}, 
\chi, \bar{\chi}, f_L(\mathbb{X}_L), 
\bar{f}_L(\bar{\mathbb{X}}_L), f_R(\mathbb{X}_R), 
\bar{f}_R(\bar{\mathbb{X}}_R)\right )~,
\eeq{gaughesoo2900mdmdnew}
where $f_L$ and $f_R$ are arbitrary functions. 
Correspondingly, for the ALP sigma-model
we have 
\beq
\int d^2 \sigma d^4\theta ~K\left ( \phi, \bar{\phi}, 
\chi, \bar{\chi}, f_L(\phi_L + \chi_L), 
\bar{f}_L(\bar{\phi}_L + \bar{\chi}_L), f_R(\phi_R 
+ \bar{\chi}_R), \bar{f}_R(\bar{\phi}_R + \chi_R)\right )~.
\eeq{gaughesoo2900mdmd}
We can study how the geometric data of the ALP 
transforms under such a transformation,
that is, how $E=\half(g+B)$ transforms; we find
$$ E~\longrightarrow M^t E M,$$
where
$$M = diag (1, 1, 1,1, \frac{\d f_L}{\d X_L}, \frac{\d \bar{f}_L}{\d \bar{X}_L}, 
\frac{\d f_R}{\d X_R}, \frac{\d \bar{f}_R}{\d \bar{X}_R}, \frac{\d f_L}{\d X_L},
 \frac{\d \bar{f}_L}{\d \bar{X}_L}, \frac{\d f_R}{\d X_R}, 
 \frac{\d \bar{f}_R}{\d \bar{X}_R}), $$\\
and we assumed that $E$ is ordered as in
(\ref{order161777}). Thus we see that the diffeomorphisms
along $\coker[J_+, J_-]$ induce transformations of the fiber
as for a vector bundle. This suggests that ALPs can be
thought of as a subbundle\footnote{For this to make sense,
we need to assume that the Poisson structure $\sigma$ is
regular.} of $TM$ such that the fibers lie along the leaves
of $\sigma$.

\section{Summary}
\label{summary}

In this work we have shown that, locally, for any
generalized K\"ahler manifold there exits an auxiliary space
(ALP), $M \times {\mathbb R}^{2d_s}$ with a particular
simple complex geometry and such that $M$ is a specific
quotient of this auxiliary space. This explains the
nonlinearities in the metric with respect to generalized
K\"ahler potential. Our construction is based on a simple
sigma-model argument which, we hope, clarifies the nature of
the semichiral fields.

On the mathematical side it would be nice to define the ALP
geometry intrinsically in geometrical terms, without any
reference to a generalized K\"ahler potential. If it is
possible, then we will have an alternative derivation of the
existence of a generalized K\"ahler potential. Another
interesting problem is to see if the ALPs can be defined
globally as a vector bundle, at least for regular
generalized K\"ahler manifolds.

For physics the ALP construction offers the opportunity to
solve some problems without using semichiral fields. For
example, the problem of a topological twist of $N=(2,2)$
sigma-model involves many auxiliary fields with the main
complication coming from semichiral sector. The chiral and
the twisted chiral sectors are relatively simple when it
comes to the topological twist. This suggests that the twist
should be performed in the ALPs and the quotient constructed
afterwards.

In \cite{Lindstrom:1994mw} it is shown how to construct hyperk\"ahler
 metrics using semichiral but no (twisted) chiral fields. These models
  should provide interesting applications of our linearization. 

Also the quotients and dualities of generalized K\"ahler
manifold can be analyzed in the ALP picture. Any isometry of
$M$ can be lifted to an isometry of the corresponding ALP
which commutes with the Kac-Moody isometries. We plan to
come back to these issues elsewhere. 
\bigskip\bigskip\bigskip\bigskip

\noindent{\bf\Large Acknowledgement}:
\bigskip

\noindent We are grateful to the 2006 Simons Workshop
for providing the stimulating atmosphere where this work was initiated.
 We thank M.T.Grisaru for discussions. 
UL supported by EU grant (Superstring theory)
MRTN-2004-512194 and VR grant 621-2003-3454.
The work of MR was supported in part by NSF grant no.~PHY-0354776
and Supplement for International Cooperation with Central and Eastern 
Euorpe PHY 0300634. The research of R.v.U. was supported by 
Czech ministry of education contract No. MSM0021622409 and 
by Kontakt grant ME649. The research of M.Z. was
supported by VR-grant 621-2004-3177. 

\appendix

\Section{The general $N=(2,2)$ Landau-Ginzburg models}
\label{a:LG}

The $N=(1,1)$ Landau-Ginzburg model is given by the following action
$$S = \int d^2\sigma\,d^2\theta~[D_+\Phi^\mu D_- \Phi^\nu E_{\mu\nu} 
(\Phi) + W(\Phi)],$$
where $E=\frac{1}{2}(g+B)$ with $H=dB$ and $W$ is an arbitrary real function. 
This action gives rise to the equation of motion
$$D_+ D_- \Phi^\lambda + \Gamma^{- 
\lambda}_{~\sigma
\nu} \,D_+\Phi^\sigma D_-\Phi^\nu - \frac{1}{2} g^{\lambda\nu} \d_\nu W =0.$$
We now consider the restrictions on $W$ that follow from
imposing invariance under additional supersymmetry 
transformations of the form \cite{Gates:1984nk}
$$\delta_2(\epsilon) \Phi^\mu
=\epsilon^+ D_+ \Phi^\nu J^\mu_{+\nu}(\Phi)
+ \epsilon^- D_- \Phi^\nu J^\mu_{-\nu}(\Phi)~ .$$
The kinetic and potential terms are invariant independently 
if the following conditions are satisfied
$$ J_\pm^t g = - gJ_\pm,~~~~~~~\nabla^{(\pm)} J_\pm=0~,~~~~~~~
J^\mu_{+\nu} \d_\mu W = \d_\nu W_\pm,$$
where $W_\pm$ are some functions. We also impose
the on-shell supersymmetry algebra. 
The commutator of two second supersymmetry 
transformations is
\ber
\nonumber[\delta_2(\epsilon_1), \delta_2(\epsilon_2)]\Phi^\mu \!\!&=&\! \!
2i \epsilon_1^+ \epsilon_2^+ \d_\+ \Phi^\lambda
( J^\mu_{+\nu}J^\nu_{+\lambda}) +2i \epsilon_1^- \epsilon_2^- \d_= 
\Phi^\lambda
( J^\mu_{-\nu}J^\nu_{-\lambda}) \\ 
\nonumber&&\!\! -\, \epsilon_1^+ \epsilon_2^+ D_+\Phi^\lambda D_+\Phi^\rho
{\cal N}^\mu_{~\lambda\rho}(J_+) - \epsilon_1^- \epsilon_2^- D_- 
\Phi^\lambda D_-\Phi^\rho
{\cal N}^\mu_{~\lambda\rho}(J_-) \nonumber \\ 
&& \!\! +\, (\epsilon_1^+ \epsilon_2^- + \epsilon^-_1 \epsilon_2^+) (J_{+\nu} 
^\mu J_{-\lambda}^\nu -
J_{-\nu}^\mu J_{+\lambda}^\nu) (D_+ D_- \Phi^\lambda + \Gamma^{- 
\lambda}_{~\sigma
\nu} \,D_+\Phi^\sigma D_-\Phi^\nu )~.\nonumber
\eer{secondsusy}
which should be 
$$[\delta_2(\epsilon_1), \delta_2(\epsilon_2)]\Phi^\mu =-2i \epsilon_1^+ 
\epsilon_2^+ \d_\+ \Phi^\mu
- 2i \epsilon_1^- \epsilon_2^- \d_= \Phi^\mu~.$$
Thus we find that $J_\pm$ are complex structures. 
However using the Landau-Ginzburg equations of 
motion the last term in the algebra 
can be canceled only upon the additional requirement that 
$$ \sigma^{\mu\nu} \d_\nu W=0,$$
where $\sigma = [J_+, J_-]g^{-1}$. Therefore we conclude
that $W$ should be a Casimir function\footnote{The Casimir
function of $\sigma$ has vanishing Poisson bracket with any
function from $C^\infty(M)$.} for the Poisson structure
$\sigma$ and moreover that $W$ is the real part
of a function holomorphic with respect to both $J_+$ and
$J_-$.

In $N=(2,2)$ language this implies that the only potential
terms possible are
$${\rm Re} \int d^2\sigma\,d\theta^+ d\theta^- {\cal W}(\phi)+
{\rm Re}\int d^2\sigma\,d\theta^+ d\bar{\theta}^- \tilde{\cal W}(\chi)~,$$
No potential is possible along
semichiral directions, \ie, along $\coker[J_+, J_-]$.
Indeed, due to the specific nature of semichiral fields, no
potential term can be written for them in $N=(2,2)$
language.

\Section{ALP generalized K\"ahler potential}
\label{a:11susy}

In this Appendix, we show that it is possible to start from
the definition of the ALP space (see section \ref{ALPS}) and
choose a generalized K\"ahler potential on the ALPs that is
invariant with respect to its Kac-Moody isometries; this
guarantees that it is possible to descend to the underlying
generalized K\"ahler geometry.

By definition, an ALP space is $M \times (V_L \oplus V_R)$
with BiLP geometry, where the vector spaces $V_{L,R}$ are
isotropic (null) vector spaces with respect to the metric on
the ALP space; they are spanned by $k_{L,R}$, the left and
right commuting Kac-Moody isometries (in the sense of
(\ref{Pdefinusiapep})), respectively. The product structure
$\Pi$ of the BiLP maps the Killing vectors $k_{L,R}$ to the
vectors $\Pi k_{L,R}$ that span the directions tangential to
$\coker[J^M_+, J^M_-]$. Because the isometries generated by
$k_{L,R}$ are Kac-Moody, they are {\it complex}. As the
isometries respect the BiLP data, we can choose a local
coordinates adapted both to the BiLP geometry (see
subsection \ref{BiLP}) and to the isometries:
$$k_L^a= \frac{\d}{\d \phi^a_L} - 
\frac{\d}{\d \chi^a_L}~,~~~~~~~~~
\bar k_L^a=\frac{\d}{\d \bar{\phi}^a_L} - 
\frac{\d}{\d \bar{\chi}^a_L}~,$$
and the right isometries as
$$ k_R^a=\frac{\d}{\d \phi^{a'}_R} - 
\frac{\d}{\d \bar{\chi}^{a'}_R}~,~~~~~~~~~
\bar k_R^a=\frac{\d}{\d \bar{\phi}^{a'}_R} - 
\frac{\d}{\d \chi^{a'}_R}~.$$
In these coordinates, the invariance of the generalized
K\"ahler potential implies that it is independent of
$\phi^a_L-\chi^a_L,\phi^{a'}_R-\bar\chi^{a'}_R$, {\it etc}.
However, we shall not use this explicit coordinate dependent
form.

Before we prove the main result, we explain the problem by
considering the analogous issue for holomorphic isometries
of a K\"ahler manifold. Consider a number of commuting
Killing vectors $k^a=k^a(z)\d+\bar k^a(\bar z)\bar\d$; these
generate an isometry if they preserve the K\"ahler potential
up to the real part of holomorphic functions $f_a$:
\beq
k^a K=f_a+\bar f_a~.
\eeq{kKisf}
As these vectors commute, we have
\beq
0=(k^a k^b-k^b k^a)K=k^a f_b+k^a\bar f_b-k^b f_a+k^b\bar f_a~.
\eeq{kkcomK}
As the functions $f_a$ are holomorphic, this implies
\beq
k^a f_b-k^b f_a=i c_{ab}~,~~~~~~k^a\bar f_b-k^b \bar f_a= -i c_{ab}~,
\eeq{kkobstruct}
where $c_{ab}$ are real constants. These constants are an
obstruction to the existence of an invariant K\"ahler
potential--no shift of $K$ by the real part of a holomorphic
function can eliminate all the $f_a$'s.

We now show that {\it no} such obstruction exists for the
(null) Kac-Moody isometries of an ALP space. We first
consider the left sector only; the right sector is
completely independent, and the same argument applies to it.
The condition for invariance of the geometry implies the
following condition on the generalized K\"ahler potential:
\beq
k_L^a K = f_a(\phi, \chi) + g_a(\phi, \bar{\chi}) + l_a(\bar{\phi}, \chi)~;
\eeq{klK}
This equation, the analog of (\ref{kKisf}), follows from
imposing invariance of the action (\ref{gaughesoo2900})
under the symmetries (\ref{symmetrieso12}); one may also
deduce this directly from the condition
(\ref{Pdefinusiapep}). Because the $k_L^a$ are null and span
an isotropic (null) space,
$$
g_{\mu\nu}k_a^\mu\bar k_b^\nu=0~,
$$
where for brevity we write $k_a\equiv k_L^a$ 
throughout the rest of this appendix.
Because of the BiLP structure of the ALP, 
this can be rewritten as:
$$
[k_a(\Pi\bar k_b)+(\Pi k_a)\bar k_b]K=0~,
$$
where $\Pi$ is the local product structure of the BiLP.
Because the vector fields $k_a$ commute and preserve 
$\Pi$, using (\ref{klK}) we find
$$
(\Pi k_a)(\bar f_b+\bar g_b+\bar l _b)+(\Pi\bar k_b)(f_a+g_a+l_a)=0~.
$$
The dependence of $f_a,g_a,l_a$ on the coordinates implies 
$$
(\Pi k_a)\bar f_b=0~,~~~~(\Pi k_a)\bar g_b=-k_a\bar g_b~,
~~~~(\Pi k_a)\bar l_b=k_a\bar l_b~,
$$
and hence we find that the null condition becomes
\beq
\mathrm{Re}(k_a\bar l_b - \bar k_a g_b)=0~.
\eeq{nullkgl}
Next we consider the condition that the 
Kac-Moody isometries commute; this gives
$$
0 = (k_a\bar k_b - \bar k_b k_a) K=2i\,
\mathrm{Im}(k_a\bar l_b - \bar k_a g_b)~,
$$
and thus we conclude that
\beq
k_a\bar l_b - \bar k_b g_a=0~.
\eeq{klkgeq}
In any contractible patch, this implies
\beq
g_a=k_a L(\phi,\bar\chi)+g_a^0(\phi)~,~~~~
l_a= k_a \bar L(\bar\phi,\chi) +l^0_a(\chi)~.
\eeq{glanswer}
If we substitute this into (\ref{klK}), we 
see that by shifting the generalized
K\"ahler potential 
$$
K\to K-(L+\bar L)~,
$$
we obtain a simple holomorphic transformation:
$$
k_a K = f_a(\phi, \chi) + g^0_a(\phi) + 
l^0_a(\chi)\equiv \tilde f_a(\phi, \chi)~.
$$
Now considering the commutator 
$$
0 =(k_a k_b-k_b k_a)K=k_a\tilde f_b -k_b\tilde f_a~;
$$
we find
$$
\tilde f_a=k_a\Lambda(\phi,\chi)~.
$$
Shifting the generalized K\"ahler potential by $\Lambda$
allows us to eliminate $\tilde f_a$, and hence find an
invariant $K$. We can proceed analogously for the right
Kac-Moody isometries. The reason that we found no possible
obstruction is that (\ref{klkgeq}), in contrast to the
K\"ahler case (\ref{kkcomK}), has only two terms.

\section{The descent from the ALPs}
In this appendix, we give details of the computations
involved in performing the quotient (\ref{djkdkwe000}) that
takes us from the ALP space down to the generalized K\"ahler
geometry. We evaluate the general formulas of 
Section \ref{quotients} for the special case when the $B$-field itself
is preserved by the Kac-Moody symmetries; in fact we have
\beq
k_{LA}^\mu E_{\mu\nu}= E_{\mu\nu}k_{RA}^\nu=0~.
\eeq{kEEkvanish}

We first consider the case without chiral or
twisted chiral fields, that is, $\ker[J_+, J_-]=\emptyset$
and the Poisson structure $\sigma$ is nondegenerate, and
hence its inverse is a symplectic form.
\subsection{Maximally symplectic case}
\label{cokernel}

Here we concentrate on the case when the whole manifold is
$\coker[J_+, J_-]$. An example of this situation is given by
a hyperk\"ahler manifold with $J_+$ and $J_-$ two different
noncommuting complex structures. Thus locally any
$d$-dimensional hyperk\"ahler metric can be thought of a
quotient metric of a $2d$-dimensional ALP space.

We start from the reduction of the $N=(2,2)$ action
\be
K\left(\mathbb{X}_L,\bar{\mathbb{X}}_L,
\mathbb{X}_R,\bar{\mathbb{X}}_R\right)
\ee\label{balalalal}
to $N=(1,1)$ components 
\be
\left(\begin{array}{cccc}
D_+X_L &
\Psi_{L+} & D_+X_R &
\Psi_{R+}\end{array}\right) E
\left(\begin{array}{c}
D_-X_L\\
\Psi_{L-}\\
D_-X_R\\
\Psi_{R-}\end{array}\right)
\ee\label{3939000}
with the auxiliary fields remaining. $E$ is given by the 
following formula 
\ber
E =
\left(\begin{array}{cccc}
0 & K_{LL} + JK_{LL}J &
JK_{LR}J & 0\\
0 & 0 & 0 & 0\\
0 & K_{RL} & 0 & 0\\
K_{RL} & JK_{RL}J & K_{RR}+JK_{RR}J & 0
\end{array}\right)
\eer{semiE}
Notice that there is actually no $\Psi_{L+}$ or $\Psi_{R-}$
in the action (just as half the the gauge fields dropped out
in (\ref{gaugeactaiioppsAAA})).

To compare to the ALP we introduce
\be
\Psi_{L\pm} = JD_\pm\Lambda_{L}~,~~~~
\Psi_{R\pm} =JD_\pm\Lambda_{R}~.
\ee
 It is sometimes convenient to rearrange rows and columns accordingly 
\ber
\left(
\begin{array}{cccc}
X_L & X_R & \Lambda_L & \Lambda_R
\end{array}\right)
\eer{foo123}
with the result that (\ref{semiE}) becomes
\be
E =
\left(\begin{array}{cccc}
0 & JK_{LR}J & K_{LL} + JK_{LL}J & 0\\
0 & 0 & K_{RL} & 0\\
0 & 0 & 0 & 0\\
K_{RL} & K_{RR}+JK_{RR}J & JK_{RL}J & 0
\end{array}\right)~.
\ee

Starting instead from the ALP we replace the 
semichiral fields by combinations of chiral 
and twisted chiral fields according to
\be
K(\phi_L+\chi_L,\bar{\phi}_L+\bar{\chi}_L,\phi_R+\bar{\chi}_R,\bar{\phi}_R+\chi_R).
\ee
When we go to $N=(1,1)$ components we treat L-fields and
R-fields differently in terms of integration by parts to
mimic the way the semichiral fields are treated. With
R-fields we integrate by parts with $D_+$ and with L-fields
we integrate by parts with $D_-$. Using a notation where the
$K_{ab}$, $K_{ab'}$ {\it etc.} entries are suppressed and
only the overall coefficients are written, the result is a
matrix with rows and columns according to
\ber
\left(\begin{array}{cccc|cccc}
\phi_L & \bar{\phi}_L & \phi_R & \bar{\phi}_R &
\chi_L & \bar{\chi}_L & \chi_R & \bar{\chi}_R
\end{array}\right)
\eer{foo321}
as
\be\label{ALPE}
\left(\begin{array}{cccc|cccc}
0 & 2 & -1 & 1 & 0 & -2 &1 & -1\\
2 & 0 & 1 & -1 & -2 & 0 & -1 & 1\\
1 & 3 & 0 & 2 &1 & -1 & 2 & 0\\
3 & 1 & 2 & 0 & -1 & 1 & 0 & 2\\
\hline
0 & 2 & -1 & 1 & 0 & -2 & 1 & -1\\
2 & 0 & 1 & -1 & -2 & 0 & -1 & 1\\
-1 & 1 & -2 & 0 & -1 & -3 & 0 & -2\\
1 & -1 & 0 & -2 & -3 & -1 & -2 & 0
\end{array}\right)~;
\ee
thus the coefficient of $K_{a_L\bar b_L}$ is $2$, the coefficient of 
$K_{a_L b_R}$ is $-1$, {\it etc}.

To go to the form we get from the semichiral reduction we identify
\ber
X_L &=& \phi_L + \chi_L,\\
X_R &=& \phi_R+\bar{\chi}_R.
\eer{foo213}
Since $\Psi_{L-} = Q_-\mathbb{X}_L$ we may identifiy
\be
\Psi_{L-} = Q_-(\phi_L+\chi_L) = JD_-(\phi_L - \chi_L)
\ee
and similarly
\be
\Psi_{+R} = JD_+(\phi_R-\bar{\chi}_R)
\ee
which leads to the definition
\ber
\Lambda_L &=& \phi_L-\chi_L\\
\Lambda_R &=& \phi_R-\bar{\chi}_R
\eer{foo312}
Rearranging rows and columns of the matrix 
(\ref{ALPE}) above accordingly, we get
\be
\left(\begin{array}{cccc|cccc}
0 & 0 & 0 & 2 & -1 & 1 & 0 & 0\\
0 & 0 & 2 & 0 & 1 & -1 & 0 & 0\\
0 & 0 & 0 & 0 & 0 & 0 & 0 & 0\\
0 & 0 & 0 & 0 & 0 & 0 & 0 & 0\\
\hline
0 & 0 & 1 & 1 & 0 & 0 & 0 & 0\\
0 & 0 & 1 & 1 & 0 & 0 & 0 & 0\\
1 & 1 & -1 & 1 & 0 & 2 & 0 & 0\\
1 & 1 & 1 & -1 & 2 & 0 & 0 & 0
\end{array}\right)
\ee
which should be compared to the ALP's $E$, 
(\ref{semiE}) with the entries evaluated. 

Finally let us write the quotient formula 
(\ref{djkdkwe000}) explicitly in these adapted coordinates. 
Using obvious matrix notation we can 
write the isometry vectors as follows
\beq
k_R = \left ( \begin{array}{c}
0\\
0\\
0\\
1\end{array}\right ),~~~~~~~~~~
k_L = \left ( \begin{array}{c}
0\\ 0\\ 1\\0
\end{array}\right )
\eeq{definaiospp}
and the connections are
\beq
\theta_L = \left ( \begin{array}{cccc}
J_s K_{LR}^{-1}J_s K_{RL} & J_sK_{LR}^{-1}C_{RR} 
& 1 & 0 \end{array} \right ),~~~~~~~~~\,
\theta^t_R = \left ( \begin{array}{c}
-C_{LL} K_{LR}^{-1}J_s \\ K_{RL}J_sK_{LR}^{-1} 
J_s\\ 0 \\ 1\end{array} \right ), 
\eeq{definconnectionslll}
which satisfy
$\theta_R k_R =1$ and $k_L^t \theta_L^t =1$.
Using $\tilde{E} = E - \theta_R^t (k_R^t E k_L) \theta_L$
we get 
\beq
\tilde{E} = \left ( \begin{array}{cccc}
C_{LL}K_{LR}^{-1}J_sK_{RL} & J_sK_{LR}J_s + 
C_{LL} K_{LR}^{-1} C_{RR} & 0 &0\\
-K_{RL}J_s K_{LR}^{-1}J_s K_{RL} & 
-K_{RL}J_s K_{LR}^{-1}C_{RR} &0 &0\\
0&0&0&0\\
0&0&0&0
\end{array}\right )~,
\eeq{finalEonslll}
which is exactly the $E$ in (\ref{E}) along $\coker[J_+,J_-]$.

\subsection{The general case}
\label{general}

We briefly discuss the general case; because they are so long,
we omit the explicit version of some formulae.

Including tourists in the ALP (\ie, other fields $\phi^{\cal A},\chi^{\cal A'}$
which do not belong to $\coker[J_+, J_-]$), the K\"ahler potential depends on
\be
\phi^{\cal A},\chi^{\cal A'},\phi_L^{A},\phi_R^{A'},\chi_L^A,\chi_R^{A'}~.
\ee
Going down to $N=(1,1)$ we find the metric and B-field. 
Remembering to treat
the left fields $\phi_L,\chi_L$ and right fields $\phi_R,\chi_R$
differently with respect to the partial integration 
we find an $E$ which, with rows and columns labelled according to
\be
\left(
\phi^{\cal A},\chi^{\cal A'},2(\phi_L+\chi_L)^A,2(\phi_R+\bar\chi_R)^{A'},
2(\phi_L-\chi_L)^A,2(\phi_R-\bar\chi_R)^{A'}
\right)~,
\ee\label{order161777}
can be written as
\be
\left(\begin{array}{cccccccccccc}
0 & 2 & 0 & -2 & 0 & 0 & -1 & 1 & 0 & 2 & 0 & 0\\
2 & 0 & -2 & 0 & 0 & 0 & 1 & -1 & 2 & 0 & 0 & 0\\
0 & 2 & 0 & -2 & 0 & 0 & -1 & 1 & 0 & 2 & 0 & 0\\
2 & 0 & -2 & 0 & 0 & 0 & 1 & -1 & 2 & 0 & 0 & 0\\
0 & 2 & 0 & -2 & 0 & 0 & -1 & 1 & 0 & 2 & 0 & 0\\
2 & 0 & -2 & 0 & 0 & 0 & 1 & -1 & 2 & 0 & 0 & 0\\
1 & 1 & -1 & -1 & 0 & 0 & 0 & 0 & 1 & 1 & 0 & 0\\
1 & 1 & -1 & -1 & 0 & 0 & 0 & 0 & 1 & 1 & 0 & 0\\
0 & 0 & 0 & 0 & 0 & 0 & 0 & 0 & 0 & 0 & 0 & 0\\
0 & 0 & 0 & 0 & 0 & 0 & 0 & 0 & 0 & 0 & 0 & 0\\
0 & 2 & 2 & 0 & 1 & 1 & 0 & 2 & -1 & 1 & 0 & 0\\
2 & 0 & 0 & 2 & 1 & 1 & 2 & 0 & 1 & -1 & 0 & 0
\end{array}\right)
\ee
From this we read off $h_{AB}$ (defined in (\ref{Pdjewke99930-0}));
\be
h_{AB} = JK_{RL}J~.
\ee
Here we use the matrix notation for the isometry vectors:
\be
k_R =\left(\begin{array}{c}
0\\ 0\\ 0 \\
0 \\ 0 \\ 1
\end{array}\right) ~,~~~~~~~~~~\,
k_L = \left(\begin{array}{c}
0 \\ 0 \\
0 \\ 0
\\ 1 \\ 0\end{array}\right)~ .
\ee
The connections are 
\be
\theta_R =\left(\begin{array}{c}
-C_{cL}K_{LR}^{-1}J \\ -C_{tL}K_{LR}^{-1}J \\ -C_{LL}K_{LR}^{-1}J \\
K_{RL}JK_{LR}^{-1}J \\ 0 \\ 1
\end{array}\right) ,~~~~~~~~~~\,
\theta_L = \left(\begin{array}{c}
JK_{LR}^{-1}C_{Rc} \\ JK_{LR}^{-1}A_{Rt} \\
JK_{LR}^{-1}JK_{RL} \\ JK_{LR}^{-1}C_{RR}
\\ 1 \\ 0\end{array}\right) ,
\ee
which when combined with (\ref{djkdkwe000}) 
gives the correct full $\tilde{E}$ on the quotient
space (\ref{E}).

\eject

\end{document}